\documentclass[twocolumn,nofootinbib,showpacs,superscriptaddress]{iopart}

\usepackage{amsfonts}
\usepackage{amssymb}
\usepackage{bm}
\usepackage{cite}
\usepackage{dcolumn}
\usepackage{graphicx}
\usepackage{graphics}
\usepackage[latin1]{inputenc}
\usepackage{latexsym}
\usepackage{rotating}
\usepackage{url}
\usepackage[colorlinks=true]{hyperref}
\usepackage{xspace} 
\usepackage[usenames]{color}

\definecolor {darkgreen}{rgb}{0.2,0.7,0.2}

\newcommand\be{\begin{equation}}
\newcommand\ba{\begin{eqnarray}}
\newcommand\ee{\end{equation}}
\newcommand\ea{\end{eqnarray}}

\newcommand\bw{\begin{widetext}}
\newcommand\ew{\end{widetext}}

\newcommand{\nn}{\nonumber}

\newcommand{\eff}{{\mbox{\tiny eff}}}
\newcommand{\Sc}{{\mbox{\tiny S}}}
\newcommand{\K}{{\mbox{\tiny K}}}
\newcommand{\SR}{{\mbox{\tiny SR}}}
\newcommand{\Hor}{{\mbox{\tiny H}}}
\newcommand{\HK}{{\mbox{\tiny H,K}}}
\newcommand{\HSR}{{\mbox{\tiny H,SR}}}
\newcommand{\Edd}{{\mbox{\tiny Edd}}}
\newcommand{\bol}{{\mbox{\tiny bol}}}
\newcommand{\red}{{\mbox{\tiny red}}}
\newcommand{\minimum}{{\mbox{\tiny min}}}
\newcommand{\maxi}{{\mbox{\tiny max}}}
\newcommand{\sph}{{\mbox{\tiny sph}}}

\newcommand{\B}{{\mbox{\tiny B}}}
\newcommand{\out}{{\mbox{\tiny out}}}
\newcommand{\tin}{{\mbox{\tiny in}}}

\newcommand{\ISCO}{{\mbox{\tiny ISCO}}}

\begin{document}
\title[Can the Slow-Rotation Approximation be used in EM Observations of BHs?]{Can the Slow-Rotation Approximation be used in Electromagnetic Observations of Black Holes?}

\author{%
Dimitry~Ayzenberg$^{1}$,
Kent~Yagi$^{1,2}$
and
Nicol\'as~Yunes$^{1}$
}

\address{$^{1}$~eXtreme Gravity Institute, Department of Physics, Montana State University, Bozeman, MT 59717, USA.}
\address{$^{2}$~Department of Physics, Princeton University, Princeton, NJ 08544, USA.}

\date{\today}

\begin{abstract} 

Future electromagnetic observations of black holes may allow us to test General Relativity in the strong-field regime. 
Such tests, however, require knowledge of rotating black hole solutions in modified gravity theories, a class of which does not admit the Kerr metric as a solution.
Several rotating black hole solutions in modified theories have only been found in the slow-rotation approximation  (i.e.~assuming the spin angular momentum is much smaller than the mass squared). 
We here investigate whether the systematic error due to the approximate nature of these black hole metrics is small enough relative to the observational error to allow their use in electromagnetic observations to constrain deviations from General Relativity.
We address this by considering whether electromagnetic observables constructed from a slow-rotation approximation to the Kerr metric can fit observables constructed from the full Kerr metric with systematic errors smaller than current observational errors. 
We focus on black hole shadow and continuum spectrum observations, as these are the least influenced by accretion disk physics, with current observational errors of about 10\%.  
We find that the fractional systematic error introduced by using a second-order, slowly rotating Kerr metric is at most 2\% for shadows created by black holes with dimensionless spins $\chi\leq0.6$.
We also find that the systematic error introduced by using the slowly rotating Kerr metric as an exact metric when constructing continuum spectrum observables is negligible for black holes with dimensionless spins of $\chi \lesssim 0.9$.
Our results suggest that the modified gravity solutions found in the slow-rotation approximation may be used to constrain realistic deviations from General Relativity with continuum spectrum and black hole shadow observations. 

\end{abstract}

\submitto{\CQG}

\maketitle

\section{Introduction}

In the next five to ten years, current and new, ground and space-based telescopes will provide unprecedented electromagnetic observations of black holes (BHs). The Event Horizon Telescope (EHT) is already taking images of the BH shadow of Sagittarius A* and the supermassive BH at the center of M87, placing some bounds on each BH's angular momentum, as well as providing evidence of their event horizons~\cite{Fish:2009va, Fish:2010wu, Broderick:2011vt, 2011ApJ...735..110B, 2009ApJ...697...45B, 2010ApJ...718..446J, 2015ApJ...807..150A, 2015ApJ...805..179B, 2014ApJ...794L..14G, 2008Natur.455...78D}. As more telescopes are added to the EHT, the quality of images of the shadows will only improve. Past telescopes, such as the Rossi X-ray Timing Explorer (RXTE), and current telescopes, such as the Chandra X-ray Observatory, XMM-Newton, and NuSTAR, have observed continuum spectra and K$\alpha$ iron line emissions from the accretion disk of stellar mass and supermassive BHs, allowing for estimates of the angular momenta of a few dozen BHs~\cite{2015PhR...548....1M}. ASTRO-H, launching this year, will provide significantly more accurate estimates of BH angular momenta, possibly doubling the current number of estimates and improving on past estimates over the next decade~\cite{2014arXiv1412.1173M}.

Electromagnetic observations of BHs are of paramount importance to learning about BH and accretion disk physics, but also, they can be excellent tools in experimental relativity. These observations can tell us about the mass and spin angular momenta of BHs, as well as about accretion disk properties, such as the temperature, accretion rate, short and long term evolution of the disk, and the presence and strength of magnetic fields~\cite{lrr-2013-1, Lasota:2015cga}. These electromagnetic observations, however, can also be used, at least in principle, to test General Relativity (GR) in the \emph{strong-field} regime~\cite{Psaltis:2008bb,2015arXiv150903884B,Johannsen:2015mdd}, i.e.~where the gravitational interaction is non-linear and the spacetime curvature is large. 

The astrophysical information in these electromagnetic observations is teased from the noise through modeling of the expected signal and its fitting to the data. In the BH shadow case, one observes the shadow cast by the BH on its accretion disk.  This shadow can be modeled through null-ray tracing on the BH spacetime. In the continuum spectrum case, one observes X-ray radiation emitted by the disk around the BH. This spectrum can be modeled as a black body with the disk described by the Novikov-Thorne~\cite{NTM} approach for geometrically-thin and optically-thick disks. With models for the observables at hand, one then fits the parameters of the model to the data, exploring the likelihood surface, for example, through Markov-Chain Monte-Carlo techniques. 

Such electromagnetic observations depend on the properties of the BH spacetime. BH shadows are sensitive to the location of the photon sphere, the surface inside which photon orbits are unstable~\cite{Claudel:2000yi}. In principle, the image is contaminated by accretion flow, but this can be removed without a model through image processing methods, such as gradient detection~\cite{Psaltis:2014mca}. Continuum spectra in thin disks are sensitive to emission near the inner edge of the disk, which can be well approximated by the innermost stable circular orbit (ISCO)~\cite{2015arXiv150903884B}. In principle, there is radiation also originating inside the ISCO, but this has been shown to be subdominant~\cite{2008ApJ...675.1048R, 2008ApJ...687L..25S, 2010MNRAS.408..752P, 2009ApJ...701.1076G, 2010ApJ...718L.117S, 2012MNRAS.424.2504Z}. 

Electromagnetic observations of BHs could then allow us, at least in principle, to determine whether BH spacetimes are truly described by the Kerr metric. The \textit{Kerr hypothesis} states that all isolated, stationary and axisymmetric astrophysical (uncharged) BHs are described by the Kerr metric, and therefore are completely determined by two parameters, their mass $m$ and (the magnitude of) their spin angular momentum $|\vec{J}|$~\cite{1975PhRvL..34..905R, 1967PhRv..164.1776I, 1968CMaPh...8..245I, 1971PhRvL..26.1344H, 1972CMaPh..25..152H, 1971PhRvL..26..331C}. The Kerr metric is the external spacetime of a Ricci-flat, stationary, and axisymmetric (uncharged) BH, which is a solution to the Einstein equations in vacuum, but it can also be a solution in certain modified gravity theories~\cite{Psaltis:2007cw}. Some modified theories of gravity, however, do not satisfy the Kerr hypothesis, and thus, electromagnetic observations of BHs could allow us to place constraints on these.

Rotating BH solutions in modified gravity theories, unfortunately, are not easy to find. Although some numerical solutions valid for arbitrary rotation are known~\cite{Kleihaus:2011tg,Kleihaus:2014lba,Kleihaus:2015aje}, most BH solutions are obtained in the slow-rotation approximation~\cite{PhysRevD.79.084043,Konno01082009,Pani:2009wy,Pani:2011gy,2012PhRvD..86d4037Y, 2014PhRvD..90d4066A, Maselli:2015tta,Maselli:2015yva, 2013PhRvL.110i1101W, 2013PhRvD..87h7504B, 2010PThPh.124..493H,Barausse:2015frm} (e.g.~in dynamical Chern-Simons (dCS) gravity~\cite{jackiw:2003:cmo,Smith:2007jm,CSreview,2012PhRvD..86d4037Y, Konno01082009, PhysRevD.79.084043,Konno:2014qua,Stein:2014xba,McNees:2015srl}). This approximation assumes the BH spin angular momentum $J$ is much smaller than its mass squared $m^{2}$, and thus, one expands the field equations in the ratio $|\vec{J}|/m^{2}$. Lacking an exact solution in these theories that is valid for all spin magnitudes, the regime of validity of these metrics is not clear. Thus, it is not clear either whether the approximate nature of these spacetimes has a significant impact on electromagnetic observables. Some observations may be less sensitive to the slow-rotation approximation than others, and whether this sensitivity matters ultimately depends on the accuracy of the observations.

The goal of this paper is to determine whether such approximate, slowly rotating BH solutions in modified gravity theories can be used to construct electromagnetic observables with which to test these theories. As a first step toward answering this question we investigate how well electromagnetic observables constructed from the slow-rotation approximation to the Kerr metric can fit observables constructed with the full Kerr solution. We specifically study BH shadow and continuum spectrum observations of BHs because they are strongly dependent on the background metric and weakly dependent on the properties of the accretion disk.

Whether the slow-rotation approximation can be used in tests of GR will depend on whether the errors introduced by this approximation are larger than other statistical, instrumental, environmental and systematic errors inherent in electromagnetic observations. Statistical error arises due to the finite accuracy and length of observation of telescopes, as well as due to covariances in the model parameters~\cite{2015arXiv150903884B}. Instrumental error is rooted, for example, in inaccuracies in the calibration of the telescopes, while environmental error is sourced, for example, by atmospheric events. The combination of statistical, instrumental and environmental error will be referred to here as \emph{observational error}, and it affects the accuracy to which BH properties can be estimated by roughly 10\%~\cite{2015PhR...548....1M,Psaltis:2014mca}. Astrophysical systematic error is sourced by our imperfect knowledge of the astrophysical model for the electromagnetic observable. Since accretion disks can be quite complicated and there are a number of models for them, there is still great uncertainty over which model(s) best describes observations~\cite{lrr-2013-1, Lasota:2015cga}. The impact of this uncertainty on the estimated model parameters is not quantitively known, and thus, we will not take it into account when comparing the slow-rotation systematic error to the observational error\footnote{Astrophysical systematic error should, in principle, be added to observational error, when comparing to the systematic error introduced by the slow-rotation approximation. Neglecting astrophysical systematics, thus, yields a conservative bound of the $|\vec{J}|/m^{2}$ range for which the slow-rotation approximation is valid.}. 

We show that the slowly rotating approximation of the Kerr metric, when appropriately resummed, is a very accurate representation of the full Kerr metric for continuum spectrum and BH shadow observations. The continuum spectrum of the slowly rotating Kerr metric agrees with the spectrum of the full Kerr metric to within our numerical accuracy with reduced $\chi^2 < 10^{-6}$ for BHs with $\vec{J}/m^2 \lesssim 0.9$. The BH shadows agree to within 2\%, with reduced $\chi^2 \lesssim 10^{-3}$ for BHs with $\vec{J}/m^2 \lesssim 0.6$. In both cases, the systematic error introduced by the approximate nature of the slowly rotating metric is much smaller than the observational error.

We also show that the maximum spin values mentioned above are not determined by the systematic error exceeding the observational error, but rather by the slowly rotating spacetime ceasing from being a physically valid metric. In the continuum spectrum case, curvature singularities appear outside of the BH event horizon when $|\vec{J}|/m^{2} \gtrsim 0.9$. In the BH shadow case, the photon sphere falls inside of the event horizon when $\vec{J}/m^2 \gtrsim 0.6$. These values of spin angular momenta, thus, provide natural cutoffs above which the approximate slowly rotating metric should no longer be used.

The remainder of this paper presents the details pertaining to these results. Section~\ref{sec:sol} presents the Kerr and slowly rotating solutions, as well as the relevant properties of each. Section~\ref{sec:spec} details how the continuum spectrum is calculated, the methodology of our analysis, and the results. Section~\ref{sec:shad} presents the same for the BH shadow.
Section~\ref{sec:disc} concludes by summarizing our results and discussing the implications. Throughout, we use the following conventions: the metric signature $(-,+,+,+)$; Latin letters in index lists stand for spacetime indices; geometric units with $G=c=1$ (e.g. $1 M_\odot$ becomes 1.477 km by multiplying by $G/c^2$ or $4.93\times10^{-6}$ s by multiplying by $G/c^3$), except where otherwise noted.

\section{Rotating BH Solutions}
\label{sec:sol}

The Kerr metric in Boyer-Lindquist coordinates ($t,r,\theta,\phi$) is given by
\begin{eqnarray}
ds_{\K}^2=&-\left(1-\frac{2mr}{\Sigma_{\K}}\right)dt^2-\frac{4mar\sin^2\theta}{\Sigma_{\K}}dtd\phi+\frac{\Sigma_{\K}}{\Delta_{\K}}dr^2
\nonumber \\
&+\Sigma_{\K} 
d\theta^2+\left(r^2+a^2+\frac{2ma^2r\sin^2\theta}{\Sigma_{\K}}\right)\sin^2\theta 
d\phi^2,
\label{eq:Kerr-metric}
\end{eqnarray}
with $\Delta_{\K}\equiv r^2-2mr+a^2$ and $\Sigma_{\K}\equiv r^2+a^2\cos^2\theta$. Here $m$ is the mass of the BH and $a\equiv J/m$ is the Kerr spin parameter, where $J:=|\vec{J}|$ is the magnitude of the BH spin angular momentum.

The slowly rotating approximation to the Kerr metric is found by expanding the Kerr metric to a given order in the dimensionless spin parameter $\chi\equiv a/m$. To quadratic order, one finds
\begin{eqnarray}
ds_{\SR}^2&=-\left(f_{\Sc} +\frac{2ma^2\cos^2\theta}{r^3}\right)dt^2-\frac{4ma\sin^2\theta}{r}dtd\phi
\nn \\
&+\left[\frac{1}{f_{\Sc}} -\frac{a^2}{r^2 f_{\Sc}^2}\left(1 - f_{\Sc} \cos^{2}\theta\right)\right]dr^2+\left(r^2+a^2\cos^2\theta\right)d\theta^2
\nonumber \\
&+\left[r^2+a^2\left(1+\frac{2m}{r}\sin^2\theta\right)\right]\sin^2\theta d\phi^2\,,
\label{eq:SR-metric}
\end{eqnarray}
where $f_{\Sc} = 1 - 2 m/r$ is the Schwarzschild factor. Note that the controlling factor in this approximation, i.e.~the background metric, is nothing but the Schwarzschild metric in Schwarzschild coordinates. Note also that both the background and its perturbation diverge at $f_{\Sc} = 0$, i.e. at the Schwarzschild event horizon due to a coordinate singularity. Note finally that in the equatorial plane ($\theta=\pi/2$) the Kerr metric and the slowly rotating metric only differ in the $g_{rr}$ component\footnote{Certain rotating BH solutions valid for arbitrary spin in modified theories of gravity, such as rotating dilaton BHs~\cite{Yazadjiev:1999ce}, share the same property.}.

This paper is concerned with studying whether the approximate slowly rotating BH solutions of certain modified gravity theories suffice to carry out tests of GR with electromagnetic observations. This is important because exact BH solutions valid for any spin are currently not available in many modified theories. To address this problem we will treat the Kerr metric of Eq.~\eref{eq:Kerr-metric} as the correct solution of Nature for rotating BHs and study whether the slowly rotating metric of Eq.~\eref{eq:SR-metric} is sufficiently accurate to model the true electromagnetic observables. That is, we will use the Kerr metric to create simulated data for electromagnetic observables and then attempt to extract these observables with a model based on the approximate, slowly rotating metric. The whole point of this analysis is to determine the range of spins for which the systematic error we will incur in the extraction of parameters from this approximate model is smaller than the statistical errors due to observational uncertainties. 

One can enhance the accuracy of the model for electromagnetic observables constructed with the slowly rotating metric by \emph{resumming} the latter in spin. An example of what we mean by resummation in this paper is the replacement $r\to \Sigma_{\K}^{1/2}$. Of course, if one does not know the functional form of the exact solution valid for all spins (i.e.~the Kerr metric in this case), one does not know how to resum the approximate solution. Indeed, there are an infinite number of possible resummations of the slowly rotating metric that generate an infinite number of different resummed metrics, all of which agree in the far-field regime but potentially disagree close to the event horizon. Perhaps, the most straightforward resummation one can think of is to simply treat the slowly rotating metric as \emph{exact} (i.e.~not expanding in $\chi$ when computing observables) and study the consequences of making such a choice. Of course, the electromagnetic model one will obtain from this resummed treatment of the slowly rotating metric will still be different from the simulated data constructed from the Kerr metric, which will thus still introduce a systematic error in the extraction of parameters. 

Resumming the slowly rotating metric is not only done to enhance the accuracy of the model, but also to simplify the calculation of electromagnetic observables. A consistent treatment in slow-rotation, where one expands all calculations to second order in $\chi$, is not straightforward when computing electromagnetic observables through \emph{numerical} ray-tracing algorithms. This is simply because once the slowly rotating metric has been numerically sampled discretely, the null geodesic equations can no longer be expanded in $\chi$. Henceforth, we treat the slowly rotating metric as exact when computing the electromagnetic model that we fit to the simulated Kerr data.

\subsection{Properties of the BH Metrics}

\subsubsection{Event Horizon}
\label{sec:EH}
The event horizon is defined as a null surface created by null geodesic generators. The normal to the surface $n^a$ must itself be null, thus event horizons must satisfy the horizon equation
\begin{equation}
g^{ab}\partial_aF\partial_bF=0,\label{horeq}
\end{equation}
where $F(x^a)$ is a level surface function that defines the location of the horizon, where $n_a=\partial_aF$. Since the spacetime is stationary, axisymmetric, and reflection symmetric about the poles and the equator, the level surfaces can only depend on radius. Without loss of generality, we let $F(x^a)=r-r_{\Hor}$, where $F=0$ yields the radial horizon location. Equation~(\ref{horeq}) then becomes $g^{rr}=0$, and this metric element in the Kerr and resummed slowly rotating BH spacetimes is given by
\begin{eqnarray}
g^{rr}_{\K}&=\frac{\Delta_{\K}}{\Sigma_{\K}} = \frac{r^2-2mr+a^2}{r^2+a^2\cos^2\theta},
\\
g^{rr}_{\SR}&=
\frac{r^3\left(1-\frac{2m}{r}\right)^{2}}{r^3-2mr^2-a^2r\sin^2\theta-2ma^2\cos^2\theta}.
\label{eq:guprrSR}
\end{eqnarray}
Notice that we do not re-expand $g^{rr}_{\SR}$ in slow-rotation, since we are treating the slowly rotating metric as exact. This is the first example we encounter in this paper that exemplifies the resummation described earlier. 

We find the horizon location by solving $g^{rr}=0$ with the Kerr and the resummed slowly rotating metric for the horizon radius, which yields
\begin{eqnarray}
r_{\HK}&=m+(m^2-a^2)^{1/2},
\\
r_{\HSR}&=2m.
\end{eqnarray}
Notice that the horizon location of the resummed slowly rotating Kerr metric agrees with that of the Schwarzschild metric, yet it is different from expanding $r_{\HK}$ to quadratic order in spin. 

\subsubsection{Conserved Quantities}
Both the Kerr metric and slowly rotating metric are stationary and axisymmetric, and thus, each possesses a timelike and an azimuthal Killing vector, implying the existence of two conserved quantities: the specific energy $E$ and the $z$-component of the specific angular momentum $L_z$. The corresponding components of the four-momentum are $p_t=-E$ and $p_\phi=L_z$. From this we find two of the geodesic equations:
\begin{eqnarray}
\dot t=&\frac{Eg_{\phi\phi}+L_zg_{t\phi}}{g_{t\phi}^2-g_{tt}g_{\phi\phi}},
\label{eq:tdot}
\\
\dot \phi=&-\frac{Eg_{t\phi}+L_zg_{tt}}{g_{t\phi}^2-g_{tt}g_{\phi\phi}},
\label{eq:phidot}
\end{eqnarray}
where the overhead dot represents a derivative with respect to the affine parameter (proper time for a massive particle). 

With these constants of the motion, we can write the motion of test particles on these backgrounds in first-order form. Substituting Eqs.~\eref{eq:tdot} and~\eref{eq:phidot} into the normalization condition $u^au_a=-1$, where $u^a=(\dot t,\dot r,\dot\theta,\dot\phi)$ is the particle's four-velocity, we find
\begin{equation}
g_{rr}\dot r^2+g_{\theta\theta}\dot \theta^2=V_\eff(r,\theta;E,L_z),
\end{equation}
where the effective potential is
\begin{equation}
V_\eff\equiv\frac{E^2g_{\phi\phi}+2EL_zg_{t\phi}+L_z^2g_{tt}}{g_{t\phi}^2-g_{tt}g_{\phi\phi}}-1.\label{Veff}
\end{equation}

We restrict attention to equatorial, circular orbits of test particles in these backgrounds. We can then solve $V_\eff=0$ and $\partial V_\eff/\partial r=0$ for $E$ and $L_z$ to find
\begin{eqnarray}
E=&-\frac{g_{tt}+g_{t\phi}\Omega}{\sqrt{-\left(g_{tt}+2g_{t\phi}\Omega+g_{\phi\phi}\Omega^2\right)}},\label{E}
\\
L_z=&\frac{g_{t\phi}+g_{\phi\phi}\Omega}{\sqrt{-\left(g_{tt}+2g_{t\phi}\Omega+g_{\phi\phi}\Omega^2\right)}},\label{Lz}
\end{eqnarray}
where
\begin{equation}
\Omega=\frac{d\phi}{dt}=\frac{-g_{t\phi,r}+\sqrt{\left(g_{t\phi,r}\right)^2-g_{tt,r}g_{\phi\phi,r}}}{g_{\phi\phi,r}},\label{Omega}
\end{equation}
represents the angular velocity of equatorial circular geodesics, i.e.~the angular velocity of zero angular-momentum observers. Since the Kerr metric and the slowly rotating metric in the equatorial plane are the same except for the $g_{rr}$ component, the energy and angular momentum are also the same in both spacetimes as neither depends on $g_{rr}$.

\subsubsection{ISCO}

The ISCO plays a very important role in the calculation of the continuum spectrum of an accretion disk model. This is because any test particle inside the ISCO plunges, rapidly crossing the event horizon. Many accretion disk models assume the inner radius of the disk is at the ISCO, an assumption well-motivated by physical arguments, simulations, and observational evidence~\cite{2008ApJ...675.1048R, 2008ApJ...687L..25S, 2010MNRAS.408..752P, 2009ApJ...701.1076G, 2010ApJ...718L.117S}. We make this same assumption throughout this work. 

The ISCO radius can be found by substituting Eqs.~(\ref{E}) and~(\ref{Lz}) into Eq.~(\ref{Veff}), and then solving $\partial^2V_\eff/\partial r^2=0$ for r. Note this will be the same in both Kerr and the slowly rotating spacetime on the equatorial plane because Eqs.~(\ref{Veff})--(\ref{Lz}) do not depend on $g_{rr}$. The ISCO radius for equatorial geodesics is then
\begin{equation}
r_{\ISCO}=m\left\{3+Z_2\mp\left[\left(3-Z_1\right)\left(3+Z_1+2Z_2\right)\right]^{1/2}\right\},
\end{equation}
where
\begin{eqnarray}
Z_1=&1+\left(1-\chi^2\right)^{1/3}\left[\left(1+\chi\right)^{1/3}+\left(1-\chi\right)^{1/3}\right],
\\
Z_2=&\left(3\chi^2+Z_1^2\right)^{1/2},
\end{eqnarray}
and $\mp$ denotes whether the accretion disk's angular momentum is in the same ($-$) or opposite ($+$) direction as the BH's angular momentum.

The ISCO of the resummed slowly rotating Kerr metric can be inside of its event horizon, since the former can shrink all the way down to $m$ for maximal spins ($\chi = 1$), while the latter stays at $2 m$. This is, of course, a problem if one is concerned with observables that depend sensitively on the location of the ISCO, as is the case for continuum spectrum observations. However, as we will discuss in the following section, there is an upper spin limit above which the resummed slowly rotating metric is no longer valid (essentially due to the metric determinant vanishing); this upper limit excludes values of spin for which the location of the ISCO is smaller than $2 m$. Thus, the fact that the ISCO can un-physically enter the horizon for the resummed slowly rotating Kerr metric will not be a concern in this paper. 

\subsubsection{Metric Determinant}
The metric determinant is necessary to calculate observables in continuum spectrum observations, as we will discuss in Sec.~\ref{sec:spec}. The metric determinants of the Kerr and the resummed slowly rotating metrics are
\begin{eqnarray}
\textbf{g}_{\K}&=-r^2,
\\
\textbf{g}_{\SR}&=-r^2\frac{\left(1-\frac{2m}{r}+\frac{a^2}{r^2}\right)\left(1-\frac{2m}{r}-\frac{a^2}{r^2}\right)}{\left(1-\frac{2m}{r}\right)^2},
\end{eqnarray}
respectively. Notice that if we were to consistently expand $\textbf{g}_{\SR}$ to ${\cal{O}}(\chi^{2})$, we would find that it equals $\textbf{g}_{\K}$ plus uncontrolled remainders of ${\cal{O}}(\chi^{4})$, but again, we do not do so since we treat the slowly rotating metric as exact.

\begin{figure}[hpt]
\begin{center}
\includegraphics[width=0.5\columnwidth{},clip=true]{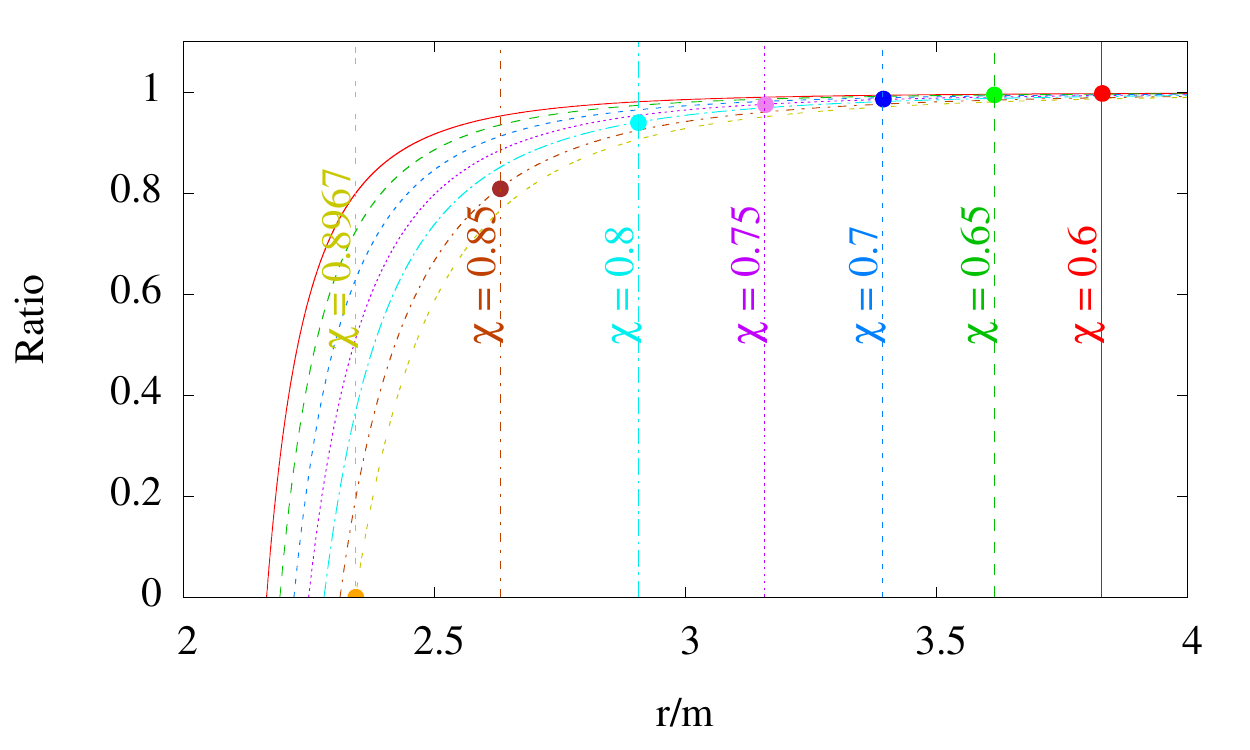}
\caption{(Color Online) Ratio between the metric determinant of the slowly rotating spacetime and the metric determinant of the Kerr spacetime. Vertical lines represent the ISCO radius. Each line style and color represents a different value of spin as labeled. Note the ratio deviates significantly from $1$ for large spin and very near the ISCO radius. The points indicate the metric determinant at the ISCO. 
\label{fig:detg}}
\end{center}
\end{figure}

The metric determinant can be used to define a region of validity for the resummed slowly rotating metric in the context of the continuum spectrum. This is because the energy flux goes as $1/\sqrt{-\textbf{g}}$, as we will discuss in Sec.~\ref{sec:spec}, and thus it diverges when the determinant vanishes. Physically, this is because the determinant of the metric is tied to the Lorentz signature of spacetime. Requiring that the determinant be strictly negative everywhere outside the ISCO requires that $\chi \lesssim 0.8967$.  Figure~{\ref{fig:detg}} shows the radial profile of the ratio between the metric determinants of the slowly rotating and Kerr spacetimes.

With this cutoff there are still values of $\chi$ for which the slowly rotating metric determinant changes sign outside the horizon radius but inside the ISCO. The Lorentz signature not being preserved outside the horizon suggests the slowly rotating solution is not well-justified for those values of spin. As we will discuss, though, the continuum spectrum calculation we use ignores any behavior below the ISCO radius; this is supported by simulations that take into account emission inside the ISCO radius and find that the contribution of that emission is below typical observational errors~\cite{2012MNRAS.424.2504Z}. Thus, continuum spectrum observables in this model should not be affected by the Lorentz signature changing sign outside the horizon, as long as it does not do so outside the ISCO.

\subsubsection{Redshift}
\label{sec:redshift}
Photons emitted in the strong-field region of a BH are greatly redshifted when they climb out of the extreme gravitational potential.  The redshift factor is defined by
\begin{equation}
g\equiv\frac{E_o}{E_e}=\frac{\left(p_a u^a\right)_o}{\left(p_a u^a\right)_e},\label{rs}
\end{equation}
where $p_a$ is the four momentum of a photon traveling from the emitter to the observer, and $u^a_o$ and $u^a_e$ are the four velocities of the observer and the emitter, respectively. Both the Kerr and the resummed slowly rotating metrics are independent of the $t$ and $\phi$ coordinates, so the corresponding components of the photon's four momentum are conserved
\begin{equation}
p_a=\left(p_t,p_r,p_\theta,p_\phi\right)=\left(-E,p_r,p_\theta,L_z\right).
\end{equation}
Treating the observer as static, $u^{a}_{0} = (1,0,0,0)$, the numerator of Eq.~(\ref{rs}) is trivially $\left(p_a u^a\right)_o=-E$. To calculate the denominator we need to know the four velocity of the emitter, that is, the orbiting emitting material. For material in a circular orbit on the equatorial plane, this is given by
\begin{equation}
u^a_{e}=u^t_{e}\left(1,0,0,\Omega\right),
\end{equation}
where
\begin{equation}
u^t_{e}=\frac{1}{\sqrt{-\left(g_{tt}+2g_{t\phi}\Omega+g_{\phi\phi}\Omega^2\right)}},\label{ut}
\end{equation}
to enforce the timelike normalization condition. The denominator of Eq.~(\ref{rs}) is now given by $\left(p_au^a\right)_e=-Eu^t_e+u^t_e\Omega L_z$, and the redshift factor is thus
\begin{equation}
g=\frac{\sqrt{-\left(g_{tt}+2g_{t\phi}\Omega+g_{\phi\phi}\Omega^2\right)}}{1-\Omega \, \xi},\label{reds}
\end{equation}
where $\xi:=L_z/E$. 

Let us now make some standard assumptions about the observer to simplify the redshift expression. We first assume the observer is at spatial infinity $(r=+\infty)$ at an inclination angle $\iota$ relative to the BH, i.e.~$\iota$ is the angle between the rotation axis of the BH and the observer's line of sight. The celestial coordinates $(\alpha,\beta)$ of the observer are defined as the apparent angular distances of the object on the celestial sphere, measured along the directions perpendicular and parallel to the rotation axis of the BH when projected onto this sphere, respectively. The celestial coordinates in terms of the photon momentum can then be written as
\begin{eqnarray}
\alpha&=\lim_{r\to\infty}\frac{-rp^{(\phi)}}{p^{(t)}},
\\
\beta&=\lim_{r\to\infty}\frac{rp^{(\theta)}}{p^{(t)}},
\end{eqnarray}
where $p^{(a)}$ denotes the components of the photon's four momentum with respect to a locally non-rotating reference frame~\cite{1972ApJ...178..347B}. $p^{(a)}$ and $p^a$ are related through a coordinate transformation (e.g. $p^\phi=p^{(\phi)}/\sin\iota$). Using the fact that the BH metrics we work with are asymptotically flat, $\alpha=-\xi/\sin\iota$. For simplicity, we will neglect light-bending in the continuum spectrum, and thus $\alpha$ can be related to the coordinates of the emitting point in the disk via $\alpha=r\cos\phi$, where $\phi=0$ corresponds to the line of nodes at which the observer's plane overlaps with the disk. Solving for $\xi$, one then finds $\xi=-r\cos\phi\sin\iota$, and substituting $\xi$ into Eq.~(\ref{reds}), one finally obtains~\cite{Moore:2015bxa}
\begin{equation}
g=\frac{\sqrt{-\left(g_{tt}+2g_{t\phi}\Omega+g_{\phi\phi}\Omega^2\right)}}{\left(1+r\Omega\cos\phi\sin\iota\right)}.\label{eq:redshift}
\end{equation}
%

\section{Continuum Spectrum}
\label{sec:spec}

We model the accretion disk with the Novikov-Thorne approach~\cite{NTM}, the standard general relativistic model for geometrically-thin and optically-thick accretion disks. This model assumes the disk is in the equatorial plane and the disk particles move on nearly geodesic circular orbits, i.e.~geodesics except for a small radial momentum. There are three equations that describe the time-averaged radial structure of the disk, two of which are used for calculating the continuum spectrum,
\begin{eqnarray}
\dot M=-2\pi\sqrt{-\textbf{g}}\Sigma(r) u^r=\mbox{constant},&
\\
\mathcal{F}(r)=\frac{\dot M}{4\pi\sqrt{-\textbf{g}}}f(r),&
\label{eq:Eflux}
\end{eqnarray}
where $\dot M$ and $\mathcal{F}$ are the time-averaged mass accretion rate and radially-dependent energy flux, respectively. $\Sigma(r)$ is the surface density, $u^r$ is the radial four-velocity of the disk particles, $\textbf{g}$ is the determinant of the near equatorial-plane metric in cylindrical coordinates, and the function $f(r)$ is defined by
\begin{equation}
f(r)=\frac{-\partial_r\Omega}{\left(E-\Omega L_z\right)^2}\int^r_{r_{\tin}}\left(E-\Omega L_z\right)\left(\partial_{r'} L_z\right)dr'.
\end{equation}
Here $r_{\tin}$ is the inner radius of the accretion disk which we choose to be the location of the ISCO. Note that the energy flux computed with the Kerr metric and the resummed slowly rotating metric only differ through the $\textbf{g}$ factor, since $E$, $L_{z}$ and $\Omega$ do not differ in these two metrics on the equatorial plane.

One could use $\dot M$ as a parameter in the continuum spectrum model, but another commonly used parameter is the Eddington ratio, $\ell=L_{\bol}/L_{\Edd}$, i.e.~the ratio between the bolometric and Eddington luminosities, where $L_{\Edd}=1.2572\times10^{38}\left(M/M_{\odot}\right)\mbox{erg/s}$. We here use this ratio to write $\dot M$ in terms of the observable $\ell$, which can be re-expressed as $\dot M/\dot M_{\Edd}$. The accretion rate $\dot M$ can then be written as
\begin{equation}
\dot M=\ell\dot M_{\Edd}.
\end{equation}
We define the radiative efficiency $\eta\equiv L_{\Edd}/\dot M_{\Edd}$, the efficiency of conversion between rest-mass and electromagnetic energy. The energy radiated by a particle falling into a BH is approximately equal to the binding energy of the ISCO~\cite{1973grav.book.....M}, so the radiative efficiency can also be written as $\eta=1-E(r_{\ISCO})$. The accretion rate $\dot M$ is then
\begin{equation}
\dot M=\frac{\ell L_{\Edd}}{1-E(r_{\ISCO})}.
\end{equation}
We model the radiation emitted by the disk as a black-body, a good assumption provided the disk is in thermal equilibrium. Using the Stefan-Boltzmann law, we can relate the radial energy flux to the radial effective temperature of the disk:
\begin{equation}
T(r)=\left(\frac{\mathcal{F}(r)}{\sigma}\right)^{1/4},
\end{equation}
where $\sigma$ is the Stefan-Boltzmann constant and where, after putting the results presented above together, the radial energy flux of Eq.~\eref{eq:Eflux} can be rewritten as
\begin{eqnarray}
\mathcal{F}(r)=&\frac{\ell L_{\Edd}}{4\pi\sqrt{-\textbf{g}}\left[1-E(r_{\ISCO})\right]}\frac{-\partial_r\Omega}{\left(E-\Omega L_z\right)^2}
\nn \\
&\times\int^r_{r_{\tin}}\left(E-\Omega L_z\right)\left(\partial_{r'} L_z\right)dr'.
\end{eqnarray}
The luminosity is then given by~\cite{Bambi:2011jq}\footnote{Strictly speaking, the inclination $\iota$ in this equation is the angle between the observer's line of sight and the direction perpendicular to the disk. However, if one assumes that the rotation axis of the BH is normal to the disk, $\iota$ is equivalent to the angle between the observer's line of sight and the BH spin direction, as introduced in Sec.~\ref{sec:redshift}.}
\begin{equation}
L(\nu)=\frac{8\pi h}{c^2}\cos\iota\int^{r_{\out}}_{r_{\tin}}\int^{2\pi}_0\frac{\nu^3\sqrt{-\textbf{g}}}{e^{\left[h\nu/gk_{\B} T(r)\right]}-1}drd\phi,
\label{eq:spectrum}
\end{equation}
where $g$ is the redshift as found in Eq~\eref{eq:redshift}, $h$ is the Planck constant, $k_{\B}$ is the Boltzmann constant, $\nu$ is the observed frequency, $r_{\out}$ is the outer radius of the disk, and we have restored the speed of light $c$. As long as the latter satisfies $r_{\out}\gg m$, the choice of $r_{\out}$ does not significantly impact our ability to compare spectra in different spacetimes~\cite{Bambi:2011jq}.

\subsection{Method}

Let us treat the Kerr metric as the correct, but unknown, description of a BH and its associated spectrum as our observation, which we shall refer to as the \textit{injected synthetic signal} or \textit{injection} for short. Let us further use the spectrum calculated with the resummed slowly rotating metric as our \emph{model} and fit it to the injection. In both cases, the spectrum is computed from Eq.~\eref{eq:spectrum}, with the integrals numerically evaluated through Simpson's rule with step sizes chosen to ensure numerical error is small. For the energy flux integration, we choose a radial step size of $\delta r=0.1m$, with a much smaller step size of $\delta r=10^{-4}m$ for $r\le r_{\ISCO}+2.5m$, as the energy flux is very steep near the ISCO radius. For the luminosity integration, we choose step sizes of $\delta r=m$ and $\delta\phi=0.1$. These choices were made after a lengthy numerical investigation to guarantee that numerical error is under control.

The parameters of the spectrum model used in this paper are $\vec{\lambda} = (m,\chi,\iota,\ell)$, namely the BH mass, its dimensionless spin, the inclination angle, and the Eddington ratio $\ell$ respectively. The BHs for which spins have been found using continuum spectrum observations have fairly well-known masses and inclination angles from other observations (e.g.~modeling of orbits using variability in electromagnetic emission~\cite{2008ApJ...679L..37L, 2014ApJ...793L..29S, 2009ApJ...701.1076G, 2010ApJ...718L.122G, 2014ApJ...784L..18M, 2006ApJ...636L.113S, 2011MNRAS.416..941S}). We thus assume \emph{a priori} that these parameters are known, and for simplicity we choose the representative values $m=10M_\odot$ and $\iota=\pi/4$; the BHs for which spins have been measured using continuum spectrum observations have masses and inclination angles in the range $6.3M_\odot\leq m\leq15.65M_\odot$ and $0.36\leq\iota\leq1.30$, respectively. Moreover, since the Eddington ratio $\ell$ is weakly correlated with the spin $\chi$ in the spectrum model, we also assume that it is known \emph{a priori} and set it to $\ell=0.1$. This is a somewhat arbitrary choice that should have little to no impact on our analysis, but about half of the BHs with estimated spins from continuum spectrum observations have an estimated $\ell\approx0.1$. This then leaves the spin $\chi$ as the only parameter of the spectrum model, for which we choose a flat prior with range $-0.899\le \chi \le 0.896$. The upper bound on this range is to avoid the divergence of the metric determinant at $\chi=0.8967$ in the resummed slowly rotating spacetime. The lower bound is to avoid a super-maximally spinning BH, i.e.~$\chi-\delta\chi=-0.999$ when $\chi=-0.899$ and $\delta\chi=0.1$, where $\delta\chi$ is the average observational error to which spins have been measured.

We fit the model to the injection by minimizing their relative $\chi^2$ over all model parameters. The reduced $\chi^2$ is defined as
\begin{equation}
\chi^2_{\red}=\frac{\chi^2}{N}=\frac{1}{N}\sum^{N}_{i=1}\left[\frac{L_{\SR}(\nu_i,\chi)-L_{\K}(\nu_i,\chi^*)}{\sigma\left(\nu_i\right)}\right]^2, \label{eq:chi-squared}
\end{equation}
where the summation is over $N$ sampling frequencies $\nu_i \in (10^{15},10^{18})$ Hz with 10 samples per decade equally spaced logarithmically. This sampling choice corresponds to that made in the observed spectra of BHs with estimated spins~\cite{2008ApJ...679L..37L, 2014ApJ...793L..29S, 2009ApJ...701.1076G, 2010ApJ...718L.122G, 2014ApJ...784L..18M, 2006ApJ...636L.113S, 2011MNRAS.416..941S}.  The function $L_{\SR}(\nu,\chi)$ is the spectrum model, which depends only on the frequency $\nu$ and on the model parameter $\chi$, while the function $L_{\K}(\nu,\chi^{*})$ is the injection, which depends only on the frequency and the injected parameters $\chi^{*}$. The value of the model parameter that minimizes the reduced $\chi^{2}$ is the best-fit model parameter. A more realistic analysis would include all parameters in the vector $\vec{\lambda}$ in a Markov-Chain Monte-Carlo exploration of the likelihood surface, but we leave this for future work.

The standard deviation of the distribution, $\sigma$, in Eq.~\eref{eq:chi-squared} is modeled via
\begin{equation}
\sigma\left(\nu_i\right)=\frac{|L_{\K}\left(\nu_i,\chi^*+\delta\chi\right)-L_{\K}\left(\nu_i,\chi^*-\delta\chi\right)|}{2}, \label{eq:err}
\end{equation}
where $\chi^*$ is the injected spin of the Kerr spectrum. $\delta\chi$ serves as a means to represent the observational error in the continuum spectrum, and thus, we choose $\delta\chi=0.1$, which is comparable to or better than the error in all current BH spin measurements using continuum spectrum observations~\cite{2015PhR...548....1M}. 

\subsection{Results}

\begin{figure*}[hpt]
\includegraphics[width=0.5\columnwidth{},clip=true]{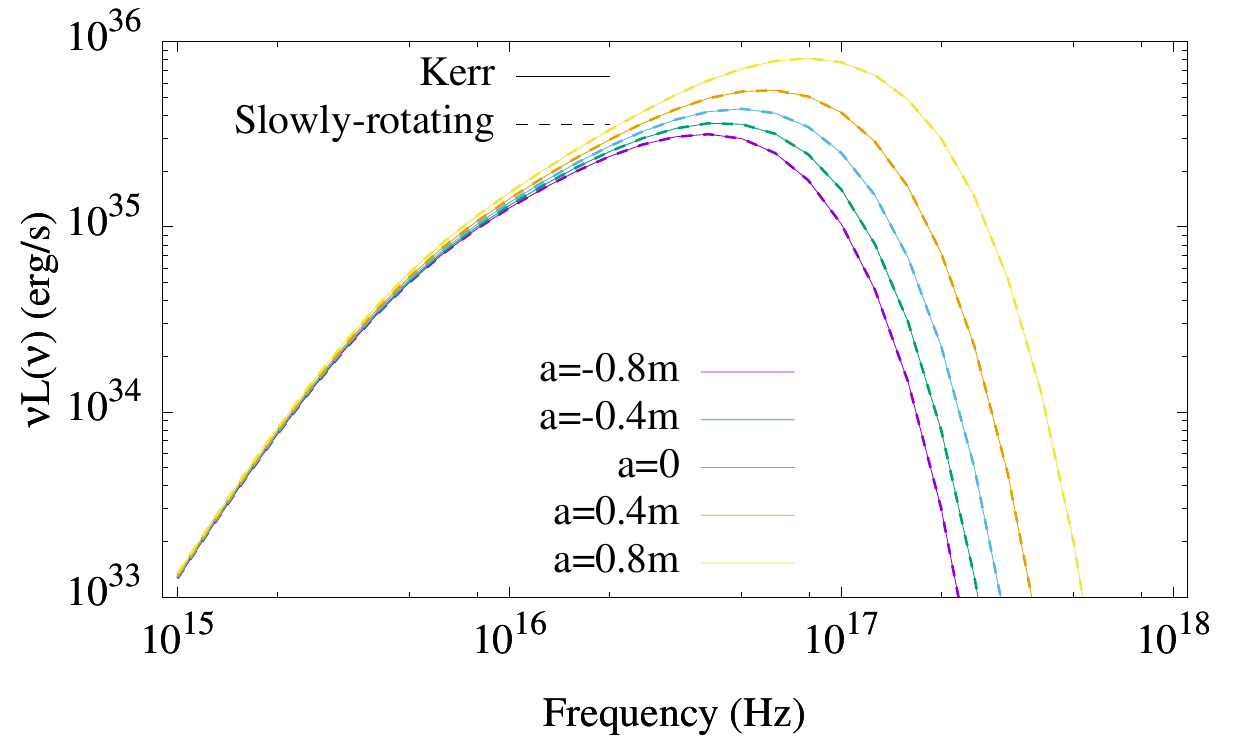}
\includegraphics[width=0.5\columnwidth{},clip=true]{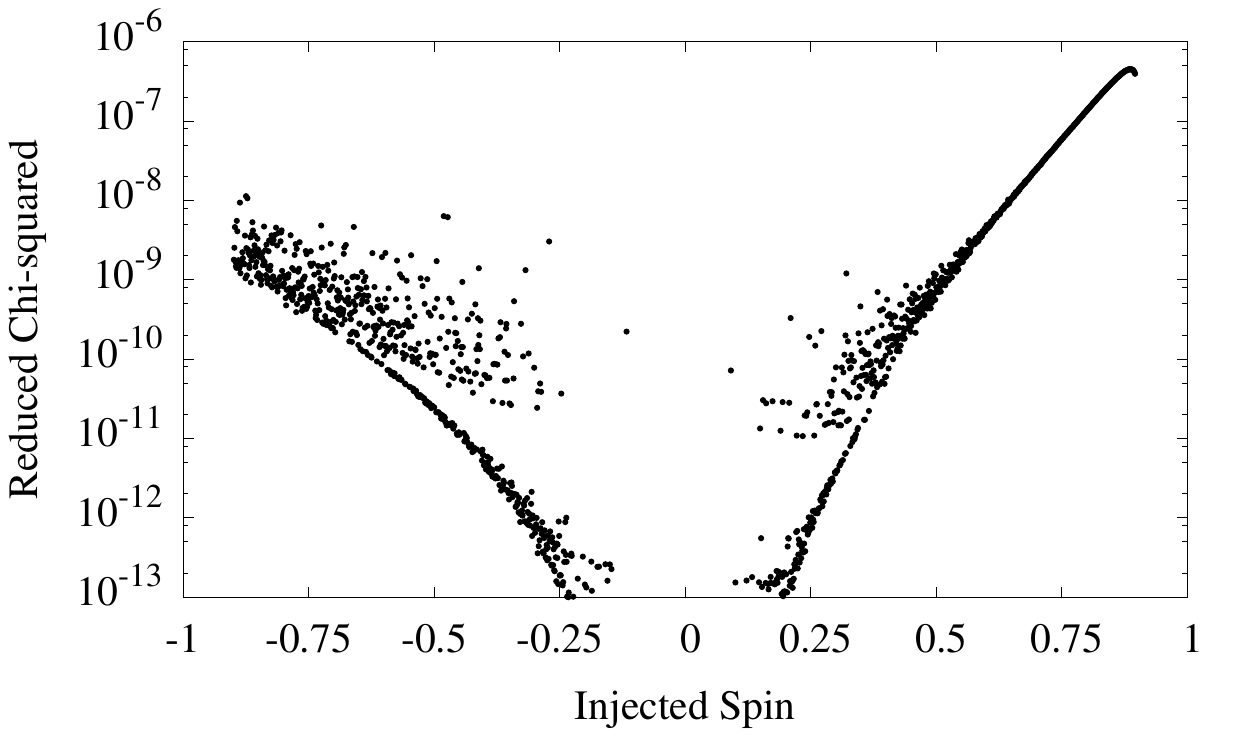}
\caption{(Color Online) Left: Continuum spectra for the Kerr and slowly rotating spacetimes for several values of spin. The Kerr spectra are shown with solid lines, while the slowly rotating spectra are shown with dashed lines. There is no obviously noticeable difference between the Kerr and slowly rotating spectra. Right: Reduced $\chi^2$ as a function of injected spin as defined in Eq.~\eref{eq:chi-squared}. The fits are very good for all values of spin, but do become worse for larger values of spin, particularly for positive spins. \label{fig:spec}}
\end{figure*}

We find that the injected spin and best fit spin agree exactly for all the values of injected spin we explored. In other words, the resummed slowly rotating spectrum and the Kerr spectrum agree with each other best when $\chi = \chi^{*}$. The left panel of Fig.~\ref{fig:spec} shows the Kerr and slowly rotating spectra for several values of spin. Observe that there is no obviously noticeable difference between the spectra in the Kerr and in the slowly rotating spacetimes. The right panel of Fig.~\ref{fig:spec} shows the minimized $\chi^2_{\red}$ as a function of the injected spin. Observe that $\chi^2_{\red} \ll 1$ for all injected spins, which shows that the resummed slowly rotating model is a very good fit of the Kerr injection. The spectra at $\chi=0$ are exactly the same, and as the spin is increased the fits get worse. The scatter in the $\chi^2_{\red}$ values is primarily due to the frequency sampling, with its range decreasing by about an order of magnitude if the discretization is made smaller by a factor of 10.

The slowly rotating spectrum can fit the Kerr spectrum very well because the two spacetimes are identical in the $\theta=\pi/2$ plane except for the $(r,r)$ component of the metrics. As mentioned previously, this component of the metric only impacts the metric determinant in the continuum spectrum calculation. As shown in Fig.~\ref{fig:detg}, the resummed slowly rotating metric determinant only shows deviations from Kerr that can be seen by eye when $r\lesssim3m$ and $\chi\gtrsim0.8$. Therefore, the fact that the model is an excellent fit to the injection is simply because the two only differ greatly when spins are very large and for trajectories very near the ISCO radius. While the accretion disk near the ISCO radius has the largest impact on the spectrum, the region where the determinants differ significantly only reaches out to about $r=r_{\ISCO}+m$ at most; this is not a large enough region to contribute significantly to the continuum spectrum observables, relative to the contributions from the rest of the accretion disk.

\section{BH Shadow}
\label{sec:shad}

Let us define the BH shadow as follows. Suppose light rays are emitted at $r=\infty$ and propagate toward the BH. If a light ray reaches an observer at $r=\infty$ after scattering, the direction of the light ray as seen by the observer is not dark. If a light ray crosses the event horizon of the BH it can never reach the observer. The light rays that do not reach the observer create a dark region, which we call the BH shadow. This is, of course, an idealization of a BH shadow; in reality, the photons that reach the observer originate from an accretion disk around the BH, which makes the shadow dependent on the properties of the accretion disk, although weakly. The shape of the boundary of the BH shadow is thus strongly dependent on the evolution of unstable spherical photon orbits near the BH, and thus, it contains information about the BH spacetime. 

The shadow boundary can be found analytically if the Hamilton-Jacobi equation is separable for the metric describing the BH, as we show explicitly in \ref{app1:kerrshadow}. Separability is not only a property of the spacetime, but  also of the coordinates chosen to describe this spacetime. For example, the Kerr metric is separable in Boyer-Lindquist coordinates, but the resummed slowly rotating metric is not. Therefore, the BH shadow boundary in the resummed slowly rotating spacetime case must be constructed numerically by solving the null geodesic equations. 

The BH shadow boundary cannot be constructed for all values of the spin of the resummed slowly rotating metric. This is simply because there is a range of spins for which the shadow boundary is inside the resummed slowly rotating horizon, i.e.~the horizon is at $r_{\HSR}=2m$, while the boundary of the shadow can be inside $2 m$ for some range of spins [see Sec.~\ref{sec:EH}]. To find these values of $\chi$,  we study the spins for which the Kerr metric has spherical photon orbits at $r=2m$, following the discussion in \ref{app1:kerrshadow}. We find that provided $\chi < 1/\sqrt{2}\approx0.707$ equatorial photon orbits exist outside of $r = 2m$ for the Kerr metric. The divergence at $r=2m$ in the slowly rotating equations of motion prevents us from doing a similar analysis for the slowly rotating metric. We thus conservatively choose to investigate resummed slowly rotating BH shadow models with $\chi\leq0.6$.

We parameterize the boundary of the BH shadow in terms of the horizontal displacement from the center of the image $D$, the average radius of the sphere $\langle R \rangle$, and the asymmetry parameter $A$. $D$, $\langle R \rangle$, and $A$ are the three BH shadow boundary observables we analyze. There are many ways to model the shape of the shadow (see e.g.~\cite{Tsukamoto:2014tja,Abdujabbarov:2015xqa}) and the conclusions of this work will be similar regardless of the chosen parameterization. The horizontal displacement $D$ is the shift of the center of the boundary of the shadow from the center of the BH and it is defined by
\begin{equation}
D\equiv\frac{|\alpha_{\minimum}-\alpha_{\maxi}|}{2},
\end{equation}
where $\alpha_{\minimum}$ and $\alpha_{\maxi}$ are the minimum and maximum horizontal coordinates of the image on the observer's viewing plane, respectively. The Kerr and the resummed slowly rotating spacetimes are axially symmetric so there is no vertical displacement of the image. The average radius $\langle R \rangle$ is the average distance of the shadow boundary from its center and it is defined by
\begin{equation}
\langle R \rangle\equiv\frac{1}{2\pi}\int^{2\pi}_{0} R(\vartheta) \; d\vartheta,
\end{equation}
where $R(\vartheta)\equiv\left[\left(\alpha-D\right)^2+\beta(\alpha)^2\right]^{1/2}$ and $\vartheta\equiv\tan^{-1}\left[\beta(\alpha)/\alpha\right]$. The asymmetry parameter is the distortion of the shadow boundary from a circle and it is defined by
\begin{equation}
A\equiv2\left[\frac{1}{2\pi}\int^{2\pi}_{0}\left(R-\langle R \rangle\right)^2d\vartheta\right]^{1/2}.
\end{equation}
%

\subsection{Method}

As with the continuum spectrum, let us treat the Kerr metric as the correct, but unknown, description of a BH and its associated BH shadow as our observation, which we shall refer to as the \textit{injected synthetic signal} or \textit{injection} for short. Let us further use the BH shadow boundary calculated with the resummed slowly rotating metric as our \emph{model} and fit it to the injection. In both cases, the BH shadow boundary is computed by first numerically evolving null geodesics with GRay~\cite{2013ApJ...777...13C}, a general relativistic ray-tracing code, and then parameterizing the boundary of the shadow as an off-centered deformed sphere~\cite{2010ApJ...718..446J}. We use GRay because the resummed slowly rotating metric is not separable, so the Hamilton-Jacobi equation must be solved numerically. 

The parameters of the BH shadow boundary model are $\vec{\lambda} = (m, \iota, \chi)$, namely the BH mass, the inclination angle and the dimensionless spin parameter respectively. We assume that the BH mass is known \emph{a priori} from other measurements (e.g.~the mass of Sagittarius A* is known to about 10\% uncertainty from observations of stellar orbits~\cite{Gillessen:2008qv, Ghez:2008ms}); for our analysis, we set $m=1$ without loss of generalization, because the BH mass only modifies the overall size of the shadow boundary without changing its shape. We further assume that the inclination is also known \textit{a priori} for simplicity, and we choose its value in a conservative fashion, as we explain below in Sec.~\ref{sec:BHshadowresults}. These priors, thus, reduce the parameter vector to just $\vec{\lambda} = (\chi)$. The prior range on $\chi$ will be chosen to be $0\leq\chi\leq0.6$, since negative values of $\chi$ would only reflect the shadow across the $\alpha$ axis. The upper bound on the prior range is due to the photon sphere falling inside of the horizon of the resummed slowly rotating metric.  

As in the spectrum case, we fit the model to the injection by minimizing their relative $\chi^{2}$ over all model parameters. The reduced $\chi^2$ in this case is defined as
\begin{equation}
\chi^2_{\red}=\frac{\chi^2}{3}=\frac{1}{3}\sum_{i=1}^3\left(\frac{\alpha_{\SR}^i(\chi)-\alpha_{\K}^i(\chi^*)}{\sigma_{i}}\right)^2,\label{chi2shad}
\end{equation}
where the three BH shadow boundary observables are $\alpha^i=[D,\langle R \rangle,A]$, $\alpha_{\K}^i$ is the injection, which depends only on the injected spin $\chi^{*}$, and $\alpha_{\SR}^i$ is the model, which depends only on the model parameter $\chi$. The value of the spin $\chi$ that minimizes the reduced $\chi^{2}$ shall be referred to as the best-fit parameter of the model. A more realistic analysis would remove the priors on the other parameters and include all parameters in a Markov-Chain Monte-Carlo exploration of the likelihood surface, but we leave this for future work. 

The standard deviations, $\sigma_{i}$, in Eq.~\eref{chi2shad} are modeled through
\begin{equation}
\sigma_{i}=\frac{\alpha_{\K}^i(\chi^*+\delta\chi)-\alpha_{\K}^i(\chi^*-\delta\chi)}{2},
\end{equation}
where $\chi^*$ is the injected spin of the Kerr shadow boundary and $\delta\chi$ represents the observational error to which this injected spin can be measured in BH shadow observations. The best BH shadow observations have only placed lower or upper bounds on spin~\cite{Doeleman19102012, 2011ApJ...735..110B}, but future observations should be able to do much better~\cite{2009ApJ...695...59D}. For simplicity, and to compare with continuum spectrum observations, we assume $\delta\chi=0.1$ below.

\subsection{Results}
\label{sec:BHshadowresults}

There are two physical parameters that determine the BH shadow boundary in the Kerr and the resummed slowly rotating spacetimes, the spin $\chi$ and the inclination angle $\iota$. We here wish to determine a \emph{conservative} bound, i.e.~a bound on how much the resummed slowly rotating shadow boundary could deviate from the Kerr shadow boundary \emph{at most}. To address this, we first fix the spin $\chi$ and find the inclination angle at which the resummed slowly rotating shadow boundary deviates the most from the Kerr shadow boundary. We then fix the inclination angle to that value and fit the slowly rotating BH shadow boundary model to the injection as a function of the model spin. This provides an upper bound on the systematic error associated with using the slowly rotating metric instead of the Kerr metric when fitting to a BH shadow observation.

\begin{figure*}[th!]
\begin{center}
\includegraphics[width=0.49\columnwidth{},clip=true]{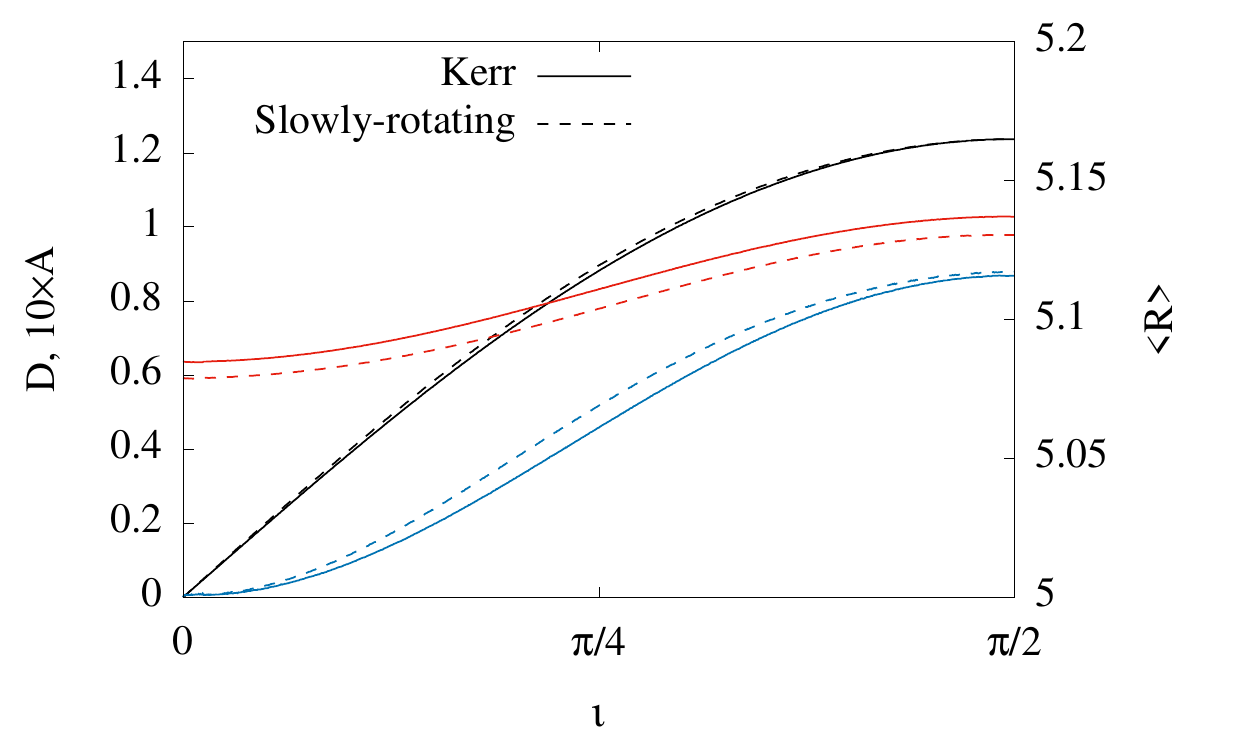}
\includegraphics[width=0.49\columnwidth{},clip=true]{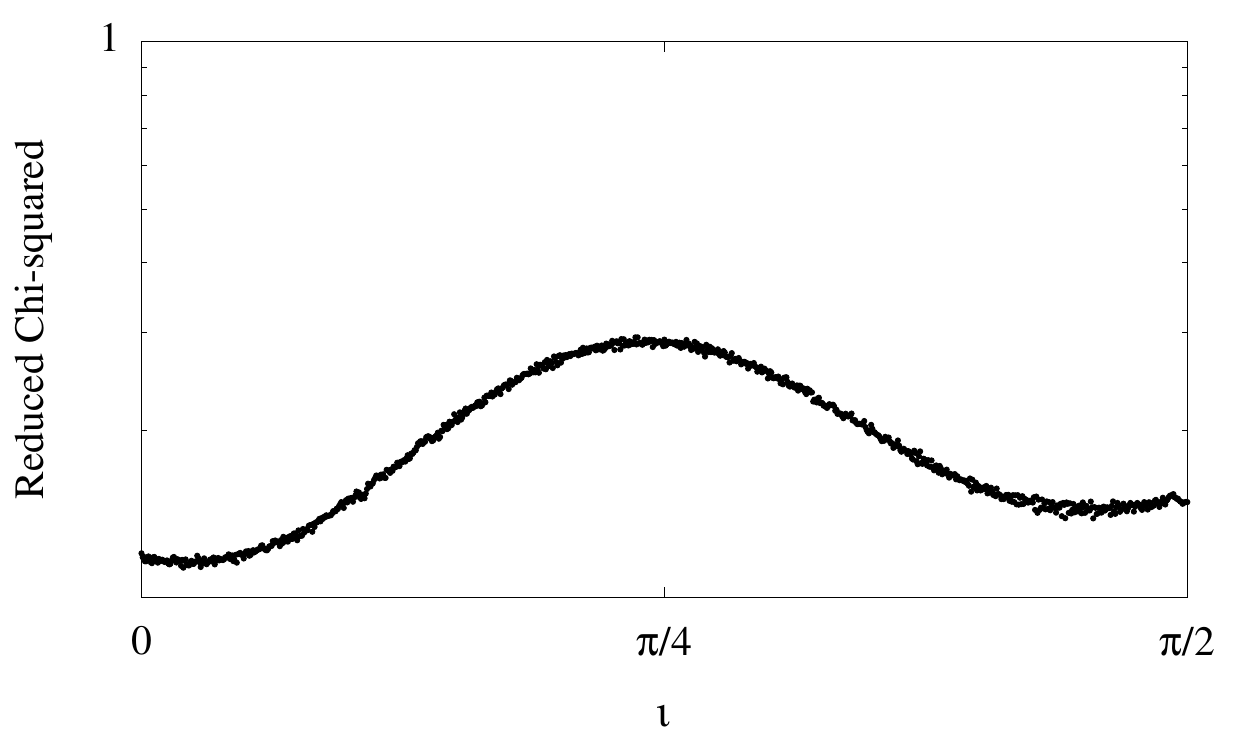}
\caption{(Color Online) Left: Displacement (black), average radius (red), and asymmetry parameter (blue) for Kerr (solid) and slowly rotating (dashed) shadows as a function of inclination angle at a spin of $\chi=0.6$. Right: $\chi^2_{\red}$ as a function of inclination angle for comparing the Kerr and slowly rotating shadow boundaries. Only shadow boundaries with equal inclination angles are compared. The spin is set at $\chi=0.6$. The errors chosen for each parameter are $(\sigma_D=0.1,\sigma_{\langle R \rangle}=0.01,\sigma_A=0.01)$. \label{fig:incprops}}
\end{center}
\end{figure*}

Let us then begin by calculating the Kerr and resummed slowly rotating BH shadow boundaries at a fixed $\chi=0.6$, but as a function of inclination angle $\iota \in (0,\pi/2)$. We parameterize the shape of each shadow boundary using the displacement, the average radius, and the asymmetry parameter, as shown in the left panel of Fig.~\ref{fig:incprops}. Observe that the difference in the average radius computed with the Kerr metric and the resummed slowly rotating is approximately constant, while the difference in the displacement and asymmetry parameters varies with inclination angle. Observe also that the maximum difference in the latter two is at $\iota\approx\pi/4$.

We confirm that the largest deviation between the shadow boundaries is near $\iota=\pi/4$ by calculating the reduced $\chi^2$ between the Kerr shadow boundary and the resummed slowly rotating shadow boundary with $\chi = 0.6$ as a function of the inclination angle. We choose the standard deviations to be an order of magnitude smaller than the injected values $(\sigma_D=0.1,\sigma_{\langle R \rangle}=0.01,\sigma_A=0.01)$. The right panel of Fig.~\ref{fig:incprops} shows that $\chi^2$ is largest at approximately $\iota=\pi/4$. Therefore, we henceforth choose $\iota=\pi/4$ for analyzing the shadow boundaries as a function of spin.

\begin{figure}[hpt]
\begin{center}
\includegraphics[width=0.5\columnwidth{},clip=true]{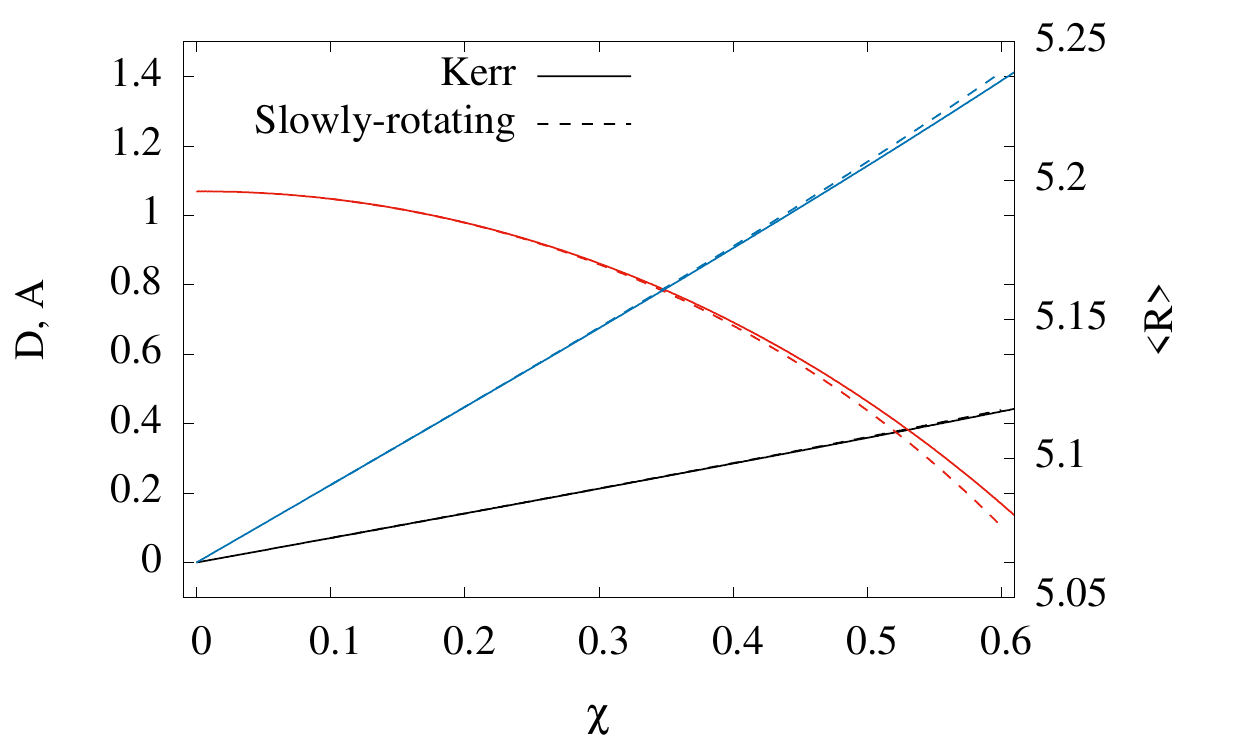}
\caption{(Color Online)  Displacement (black), average radius (red), and asymmetry parameter (blue) for Kerr (solid) and slowly rotating (dashed) shadow boundaries as a function of spin at an inclination angle of $\iota=\pi/4$.
\label{fig:spinshad}}
\end{center}
\end{figure}

Let us now fit a BH shadow model constructed with the resummed slowly rotating metric to the Kerr BH shadow injection as a function of model parameter $\chi$ with $\iota$ fixed at $\pi/4$.  As in the continuum spectrum case, we find that the injected and best fit spins agree exactly for all the values of injected spin we explored, i.e.~the resummed slowly rotating BH shadow and the Kerr shadow agree with each other best when $\chi = \chi^{*}$. Figure~\ref{fig:spinshad} shows the BH shape observables computed with the resummed slowly rotating metric and the Kerr metric using $\iota = \pi/4$ and the same values of spin. Once again, observe that the resummed slowly rotating shadow with $\chi = \chi^{*}$ matches the full Kerr shadow shadow quite well; the difference between the shadow boundaries increases, becoming noticeable by eye only at very high spin.

\begin{figure*}[htb]
\includegraphics[width=0.5\columnwidth{},clip=true]{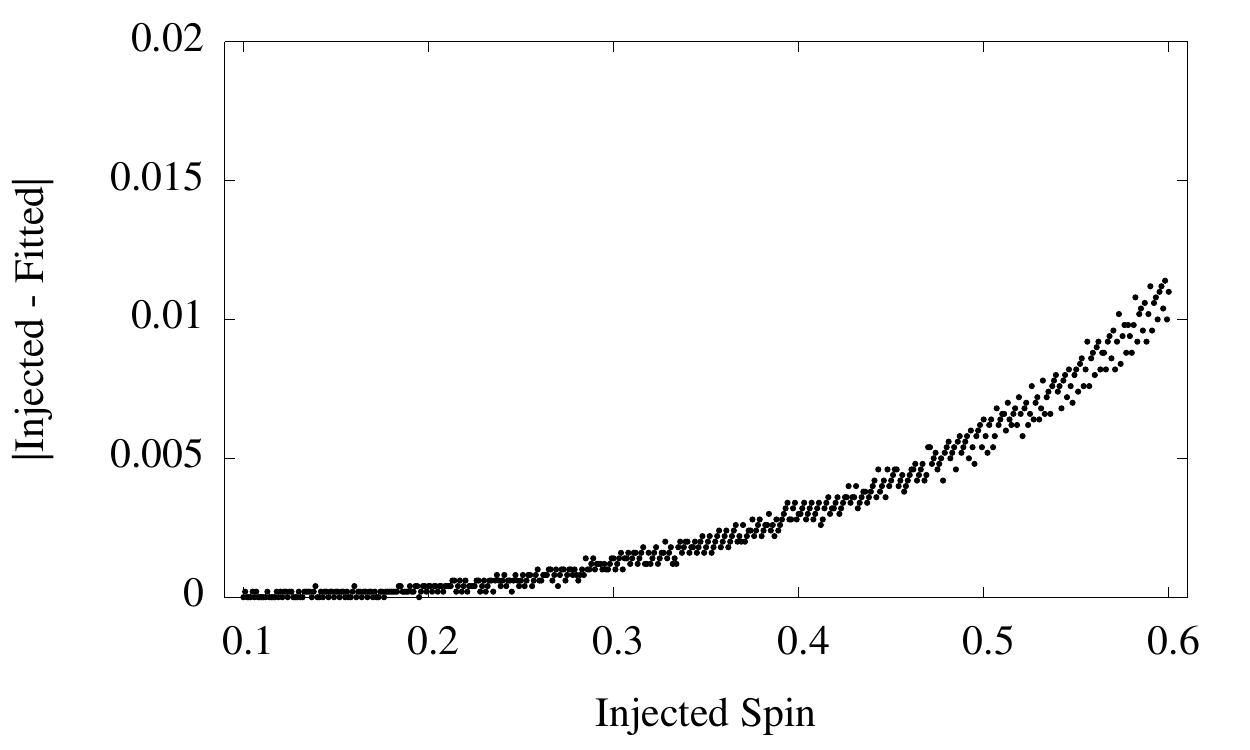}
\includegraphics[width=0.5\columnwidth{},clip=true]{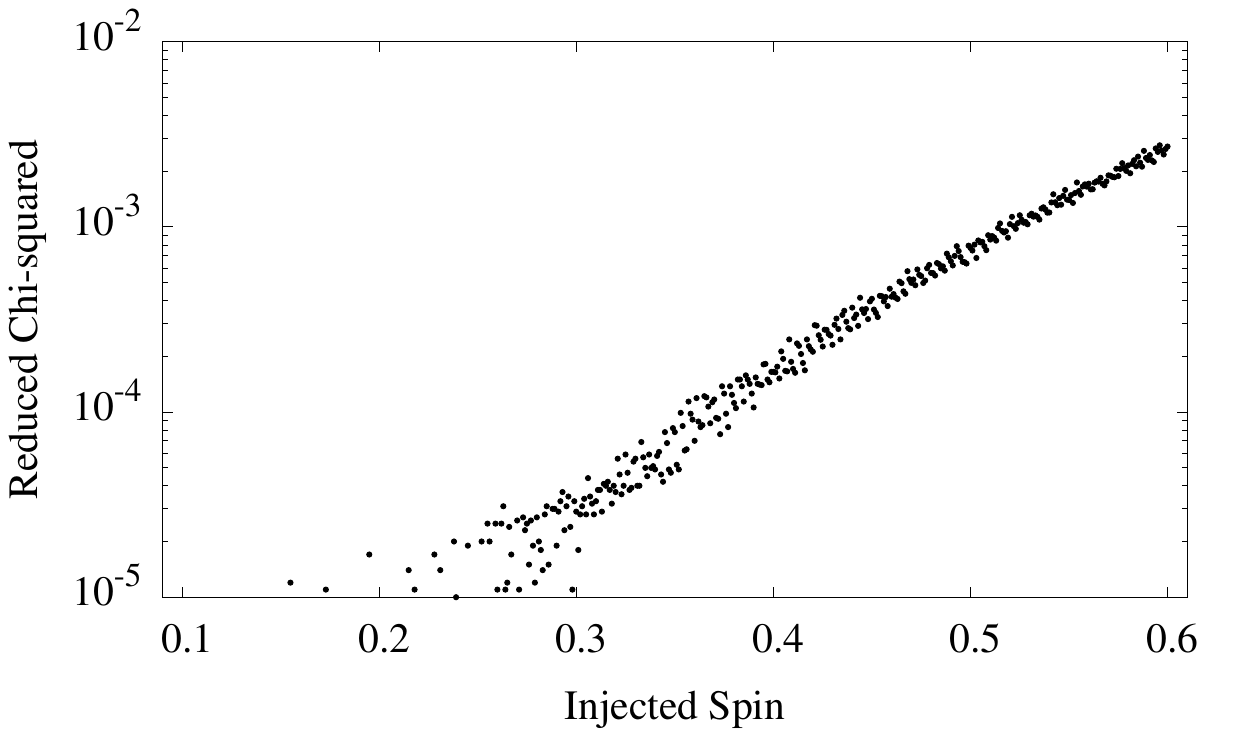}
\caption{Left: Absolute difference between the injected spin of the Kerr shadow boundary and the spin of the best fit slowly rotating shadow boundary as a function of injected spin. Right: $\chi^2_{\red}$ as defined in Eq.~\eref{chi2shad}. Inclination angle is set at $\iota=\pi/4$. \label{fig:spinfit}}
\end{figure*}

The left panel of Fig.~\ref{fig:spinfit} shows the absolute difference between the injected spin and the best fit spin as a function of injected spin. Observe that the absolute difference is at most $0.012$ at an injected spin of $0.6$, or equivalently the fractional systematic error on the estimated spin is at most 2\%. The right panel of Fig.~\ref{fig:spinfit} shows $\chi^2_{\red}$ as a function of injected spin. Observe that the goodness-of-fit deteriorates with increasing injected spin, with $\chi^2_{\red}$ reaching at most about $0.003$ at the edge of the prior range. The $\chi^2_{\red}$ at low spins is dominated by numerical errors and thus exhibits some scattering. In conclusion, the best fit spin value deviates more and more from the injected spin as the latter increases, while the fit deteriorates; however, the fractional systematic error in the estimated spin is smaller than reasonable estimates of the observational error expected in BH shadow observations~\cite{Psaltis:2014mca}.

%

\section{Discussion}
\label{sec:disc}

We have studied whether the slow-rotation approximation when constructing BH spacetimes is appropriate for modeling two electromagnetic observables associated with BH accretion disks: the continuum spectrum and BH shadows. We have found that this approximation does not introduce a significant systematic error when fitting resummed slowly rotating models to an exact Kerr injection. In the continuum spectrum case, the slowly rotating model can capture the Kerr spectrum to within our numerical accuracy. In the BH shadow case, the slowly rotating model incurs a fractional systematic error in the estimation of the spin parameter of 2\% at worst. These systematic errors are smaller than the observational errors (statistical, astrophysical and instrumental) associated with continuum spectrum and BH shadow observations~\cite{2015PhR...548....1M,Psaltis:2014mca} for all spins considered for which the slowly rotating metric is physically valid. 

We have also found that the slow-rotation approximation cannot be used to model BHs with too high a spin. In particular, one should not construct slowly rotating continuum spectra and BH shadows to model BHs with dimensionless spin larger than approximately $0.9$ and $0.6$ respectively. This is because above these values of spin the approximate spectrum and the shadow are unphysical due to the volume element vanishing and the photon sphere falling inside of the event horizon of the slowly rotating metric respectively.  

Our results suggest that the continuum spectrum and BH shadow observations can be analyzed with approximate slowly rotating BH solutions, since this approximation introduces less systematic error than the observational error of current and future telescopes. This opens the door to carrying out tests of GR with approximate, slowly rotating BH solutions in modified gravity theories using electromagnetic observations; this is important because there is a wide class of modified gravity theories, such as dCS gravity~\cite{2012PhRvD..86d4037Y, Konno01082009, PhysRevD.79.084043}, for which an exact BH solution valid for all spins is currently not known. 

The conclusions derived here, however, are only valid provided the difference between slowly rotating BHs and exact BHs in modified theories can be captured by the difference between the slowly rotating Kerr metric and the Kerr metric. This assumption could be verified by carrying out a similar study in modified gravity theories for which exact rotating BH solutions are known. Analytic and exact rotating BH solutions have not been found in many modified gravity theories (e.g. dCS gravity~\cite{2012PhRvD..86d4037Y, Konno01082009, PhysRevD.79.084043}), but there are some for which analytic, exact rotating BH solutions exist (e.g.~massive gravity~\cite{Babichev:2014tfa}). 

Yet another possible extension is to improve on the accretion disk model used in the continuum spectrum calculation. The Novikov-Thorne model (and models that are adapted from it) is the most commonly used framework when fitting for continuum spectrum observations. However, there are other models that incorporate different assumptions or different physics~\cite{lrr-2013-1}, such as the advection-dominated accretion flow (ADAF) model, which could potentially produce a significantly different continuum spectrum~\cite{2011ApJ...729...10X}. It is possible that the agreement in the continuum spectrum between the slowly rotating metric and the Kerr metric disappears in other accretion disk models and this is worth investigating in the future.

A final extension of our work could be to relax other assumptions we made in this paper to further verify the conclusions we arrived at. The resummation used for the slowly rotating metric is the simplest choice possible, but is certainly not the only one. The effect of light-bending was neglected for the continuum spectrum calculation and could be incorporated with a ray-tracing algorithm. Values of mass, inclination angle, Eddington ratio, and outer disk radius were assumed known \emph{a priori} for the continuum spectrum and BH shadow. A full multi-dimensional Bayesian analysis could be performed to see whether covariances introduced by these additional parameters have an impact on our conclusions.

\ack

D.A.~and N.Y.~acknowledge support from the NSF CAREER Grant PHY-1250636. K.Y.~acknowledges support from JSPS Postdoctoral Fellowships for Research Abroad and NSF grant PHY-1305682. Some calculations used the computer algebra-systems MAPLE, in combination with the GRTENSORII package~\cite{grtensor}.

\appendix
\section{BH Shadow in the Kerr spacetime}
\label{app1:kerrshadow}

The boundary of the BH shadow is described by the separatrix between the photon geodesics that make it out to spatial infinity and those that fall into the BH event horizon~\cite{Claudel:2000yi}. The construction of the BH shadow thus requires the solution to the null-geodesic equations, which can be derived, for example, through the Hamilton-Jacobi equation. For generic spacetimes, the null geodesic equations cannot be solved through separation of variables and must thus be treated numerically. For separable spacetimes, however, like that described by the Kerr metric in Boyer-Lindquist coordinates, the separatrix between outgoing and ingoing null geodesics can be found analytically. We here show how this is done in the Kerr spacetime in Boyer-Lindquist coordinates.

We begin with the Hamilton-Jacobi equation
\begin{equation}
\frac{\partial S}{\partial \lambda}=\frac{1}{2}g^{ab}\frac{\partial S}{\partial x^a}\frac{\partial S}{\partial x^b},\label{HJE}
\end{equation}
where $S$ is the Jacobi action, $\lambda$ is the affine parameter, and $x^a$ are generalized coordinates. If we assume separability and note that we only care about null geodesics, the Jacobi action can be written as
\begin{equation}
S=-Et+L_z\phi+S_r(r)+S_\theta(\theta).
\end{equation}
Inserting this ansatz into Eq.~(\ref{HJE}), we find the partial differential equation
\begin{eqnarray}
2\frac{\partial S}{\partial\lambda}=0=&g^{tt}E^2-2g^{t\phi}EL_z+g^{\phi\phi}L_z^2
\nn \\
&+g^{rr}\left(\frac{dS_r}{dr}\right)^2+g^{\theta\theta}\left(\frac{dS_\theta}{d\theta}\right)^2\,,
\end{eqnarray}
which through separation of variables becomes
\begin{eqnarray}
\Delta_{\K}\left(\frac{dS_r}{dr}\right)^2=\frac{1}{\Delta_{\K}}\left[E\left(r^2+a^2\right)-aL_z\right]^2-\left(L_z-aE\right)^2-\mathcal{Q}&,\label{dSr}
\\
\left(\frac{dS_\theta}{d\theta}\right)^2=\mathcal{Q}+\cos^2\theta\left[a^2E^2-\frac{L_z^2}{\sin^2\theta}\right],&\label{dSth}
\end{eqnarray}
where $\mathcal{Q}$ is the Carter constant.

The null-geodesic equations for the $r(\lambda)$ and $\theta(\lambda)$ components of the null trajectories can be found by noting that $dS/dr=p_r=g_{rr} (dr/d\lambda)$ and $dS/d\theta=p_\theta=g_{\theta\theta} (d\theta/d\lambda)$. The equations are then simply
\begin{eqnarray}
\Sigma_{\K}\frac{dr}{d\lambda}&=\pm\sqrt{\mathcal{R}}, \label{eq:drdlambda}
\\
\Sigma_{\K}\frac{d\theta}{d\lambda}&=\pm\sqrt{\Theta}\,, \label{eq:dthetadlambda}
\end{eqnarray}
where we have defined the two functions
\begin{eqnarray}
\mathcal{R}(r)&\equiv\left[E\left(r^2+a^2\right)-aL_z\right]^2-\Delta_{\K}\left[\mathcal{Q}+\left(L_z-aE\right)^2\right],\label{eqR}
\\
\Theta(\theta)&\equiv\mathcal{Q}+\cos^2\theta\left[a^2E^2-\frac{L_z^2}{\sin^2\theta}\right].
\label{eq:Theta}
\end{eqnarray}

Unstable spherical photon orbits are defined by the conditions
\begin{eqnarray}
\mathcal{R}=0, \quad
\frac{d\mathcal{R}}{dr}=0,  
\label{eq:spherical-photon-orbit0}
\\
\Theta\geq0.
\label{eq:spherical-photon-orbit}
\end{eqnarray}
For simplicity, we define the conserved quantities $\xi\equiv L_z/E$ and $\eta\equiv Q/E^2$, and solve for each using Eq.~(\ref{eqR}) and Eq.~\eref{eq:spherical-photon-orbit0} to find
\begin{eqnarray}
\xi_{\sph}&=\frac{r_{\sph}^2+a^2}{a}-\frac{2\Delta_{\K} r_{\sph}}{a\left(r_{\sph}-m\right)},\label{xisol}
\\
\eta_{\sph}&=-\frac{r_{\sph}^3\left[r_{\sph}\left(r_{\sph}-3m\right)^2-4a^2m\right]}{a^2\left(r_{\sph}-m\right)^2},\label{etasol}
\end{eqnarray}
where $r_{\sph}$ is the constant radius of the unstable spherical orbits. This radius is  constrained by the condition in Eq.~\eref{eq:spherical-photon-orbit}, which we can simplify by rewriting $\Theta$ in terms of $\xi$ and $\eta$ as
\begin{eqnarray}
\frac{\Theta}{E^2}&=\mathcal{J}-\left(a\sin\theta-\xi\csc\theta\right)^2,
\end{eqnarray}
where
\begin{equation}
\mathcal{J}=\eta+\left(a-\xi\right)^2.
\end{equation}
Therefore, the $\Theta\geq0$ condition for unstable spherical orbits implies the necessary (but not sufficient) condition $\mathcal{J}\geq0$. Substituting Eqs.~(\ref{xisol}) and~(\ref{etasol}) into the above gives the condition
\begin{equation}
\mathcal{J}=\frac{4r_{\sph}^2\Delta_{\K}}{\left(r_{\sph}-m\right)^2} \geq 0\,,
\end{equation}
which reduces simply to $\Delta_{\K}\geq0$ or $r_{\sph}\geq m+\sqrt{m^2-a^2}$, which is the Kerr horizon radius.

The BH shadow boundary is defined as the sky projection of the photon sphere as observed at spatial infinity. The conserved parameters $\xi$ and $\eta$ can be related to the celestial coordinates of the observer at infinity via
\begin{eqnarray}
\alpha&=\lim_{r\rightarrow\infty}\frac{-rp^{(\phi)}}{p^{(t)}}=-\frac{\xi_{\sph}}{\sin \iota},
\\
\beta&=\lim_{r\rightarrow\infty}\frac{rp^{(\theta)}}{p^{(t)}}=\left(\eta_{\sph}+a^2\cos^2 \iota-\xi_{\sph}^2\cot^2 \iota\right)^{1/2}.
\end{eqnarray} 
The separatrix between ingoing and outgoing photon geodesics, what we call the boundary of the BH shadow, can then be constructed by plotting $(\alpha,\beta)$ by varying $r_{\sph}$ from $r = m + \sqrt{m^{2} - a^{2}}$ to $r=4.5 m$; the latter is the photon sphere radius for a Schwarzschild BH and also the largest radius for which closed photon orbits are possible for all values of spin.

\section*{References}
\bibliography{biblio}
\bibliographystyle{iopart-num}

\end{document}